\documentclass[pra,aps,twocolumn,eqsecnum,showpacs]{revtex4}

\usepackage{dcolumn}
\usepackage{amsmath}
\usepackage{graphicx}

\begin{document}

\title{Cooperative light scattering on an atomic system with\\
degenerate structure of the ground state}

\author{A.S. Sheremet${}^1$, A.D. Manukhova${}^2$, N.V. Larionov${}^1$, D.V. Kupriyanov${}^1$}

\address{${}^1$Department of Theoretical Physics, St-Petersburg State
Polytechnic University, 195251, St.-Petersburg, Russia\\
${}^2$Department of Physics, St-Petersburg State University, 198504, St-Petersburg, Russia}
\email{kupr@dk11578.spb.edu}

\date{\today}

\begin{abstract}
We present here a microscopic analysis of the cooperative light scattering on
an atomic system consisting of $\Lambda$-type configured atoms with the spin-degenerate ground state. The results are compared with a similar system
consisting of standard "two-level" atoms of the Dicke model. We discuss
advantages of the considered system in context of its possible implications for
light storage in a macroscopic ensemble of dense and ultracold atoms.
\end{abstract}



\maketitle

\section{Introduction}
 A significant range of studies of ultracold atomic systems have focused on their complex quantum behavior in various interaction processes. Among these, special attention has been payed to the quantum interface between light and matter, and quantum memory in particular \cite{PSH,Simon,SpecialIssueJPhysB}. Most of the schemes for light storage in atomic ensembles are based on idea of the $\Lambda$-type conversion of a signal pulse into the long-lived spin coherence of the atomic ground state. The electromagnetically induced transparency (EIT) protocol in a warm atomic ensemble was successfully demonstrated in Ref. \cite{NGPSLW}, and also in Ref. \cite{CDLK}, where a single photon entangled state was stored in two ensembles of cold atoms with an efficiency of 17\%. Recent experiments on conversion of a spin polariton mode into a cavity mode with efficiency close to 90\% \cite{STTV} and on the narrow-bandwidth biphoton preparation in a double $\Lambda$-system under EIT conditions \cite{DKBYH} show promising potential for developing a quantum interface between light and atomic systems.  However, further improvement of atomic memory efficiencies is a challenging and not straightforward experimental task. In the case of warm atomic vapors, any increase of the sample optical depth meets a serious barrier for the EIT effect because of the rather complicated and mainly negative rule of atomic motion and Doppler broadening, which manifest in destructive interference among the different hyperfine transitions of alkali-metal atoms \cite{MSLOFSBKLG}. In the case of ultracold and dilute atomic gas, which can be prepared in a magneto-optical trap (MOT), for some experimental designs optical depths around hundreds are feasible \cite{FGCK}, but there are a certain challenges in accumulating so many atoms and making such a system controllable. One possible solution requires special arrangements for effective light storage in MOT in a diffusion regime; see Ref. \cite{GSOH}.

Recent progress in experimental studies of light localization phenomenon in the dense and strongly disordered atomic systems \cite{Kaiser,BHSK} encourages us to think that the storage protocols for light could be organized more effectively if atoms interacted with the field cooperatively in the dense configuration. If an atomic cloud contains more than one atom in the volume scaled by the radiation wavelength, the essential optical thickness can be attained for a smaller number of atoms than it is typically needed in dilute configuration. In the present paper we address the problem of light scattering by such an atomic system, which has intrinsically cooperative behavior. Although the problem of cooperative or dependent light scattering and super-radiance phenomenon have been well established in atomic physics and quantum optics for decades (see Refs. \cite{BEMST,Akkermns}), microscopic analysis for the atoms with degenerate ground states is still quite poorly performed in the literature \cite{Grubellier}. The microscopic calculations reported so far have been done mostly for "two-level" atoms and were basically motivated by the problem of mesoscopic description of the light transport through disordered media and by an Anderson-type localization, where transition from weak to strong disorder plays a crucial role; see  Refs. \cite{RMO,GeroAkkermns,SKKH}.

In this paper we develop a microscopic theory of the cooperative light
scattering from an atomic system consisted of $\Lambda$-type configured
atoms with the spin-degenerate ground state. The results are compared with
a similar system of "two-level" atoms of the Dicke model. We discuss
advantages of the considered system in the context of its possible implications for
the problem of light storage in a macroscopic ensemble of dense and ultracold
atoms.

\section{Theoretical framework}
\subsection{Transition amplitude and the scattering cross section}
The quantum-posed description of the photon scattering problem is based on the
formalism of $T$ matrix, which is defined by
\begin{equation}
\hat{T}(E)=\hat{V}+\hat{V}\frac{1}{E-\hat{H}}\hat{V},%
\label{2.1}%
\end{equation}
where $\hat{H}$ is the total Hamiltonian consisting of the nonperturbed part
$\hat{H}_0$ and an interaction term $\hat{V}$ such that
$\hat{H}=\hat{H}_0+\hat{V}$. The energy argument $E$ is an arbitrary complex
parameter in Eq.(\ref{2.1}). Then the scattering process, evolving from initial
state $|i\rangle$ to the final state $|f\rangle$, is expressed by the following
relation between the differential cross section and the transition amplitude,
given by the relevant $T$-matrix element considered as a function of the initial
energy $E_i$:
\begin{equation}
d\sigma_{i\to f}=\frac{{\cal V}^2}{\hbar^2
c^4}\frac{\omega'^2}{(2\pi)^2}%
\left|T_{g'\mathbf{e}'\mathbf{k}',g\,\mathbf{e\,k}}(E_i+i0)\right|^2d\Omega%
\label{2.2}%
\end{equation}
Here the initial state $|i\rangle$ is specified by the incoming photon's wave
vector $\mathbf{k}$, frequency $\omega\equiv\omega_k=c\,k$, and polarization
vector $\mathbf{e}$, and the atomic system populates a particular ground state
$|g\rangle$. The final state $|f\rangle$ is specified by a similar set of the
quantum numbers, which are additionally upscribed by the prime sign, and the
solid angle $\Omega$ is directed along the wavevector of the outgoing photon
$\mathbf{k}'$. The presence of quantization volume ${\cal V}$ in this
expression is caused by the second quantized structure of the interaction
operators; see below. The scattering process conserves the energy of input and
output channels, such that $E_i=E_f$.

Our description of interaction process of the electromagnetic field with an
atomic system is performed in the dipole approximation. This states that
the original Hamiltonian, introduced in the Coulomb gauge and valid for any
neutral charge system, has been unitarily transformed to the dipole-type
interaction with the assumption that atomic size is much smaller than a typical
wavelength of the field modes actually contributing to the interaction
dynamics. Such a long-wavelength dipole approximation see Ref. \cite{ChTnDRGr} for
derivation details leads to the following interaction Hamiltonian for an
atomic ensemble consisting of $N$ dipole-type scatterers interacting with the
quantized electromagnetic field:
\begin{eqnarray}
\hat{V}&=&-\sum_{a=1}^{N}%
\hat{\mathbf{d}}^{(a)}\hat{\mathbf{E}}(\mathbf{r}_a)+\hat{H}_{\mathrm{self}},%
\nonumber\\%
\hat{H}_{\mathrm{self}}&=&\sum_{a=1}^{N}\frac{2\pi}{{\cal V}}\sum_{s}\left(\mathbf{e}_s\hat{\mathbf{d}}^{(a)}\right)^2%
\label{2.3}%
\end{eqnarray}
The first and most important term is normally interpreted as interaction of an
$a$th atomic dipole $\mathbf{d}^{(a)}$ with electric field
$\hat{\mathbf{E}}(\mathbf{r})$ at the point of dipole location. However,
strictly defined in the dipole gauge, the latter quantity performs the
microscopic displacement field, which can be expressed by a standard expansion
in the basis of plane waves $s\equiv{\mathbf{k},\alpha}$ (where $\alpha=1,2$
numerates two orthogonal transverse polarization vectors
$\mathbf{e}_s\equiv\mathbf{e}_{\mathbf{k}\alpha}$ for each $\mathbf{k}$)
\begin{eqnarray}
\lefteqn{\hat{\mathbf{E}}(\mathbf{r})\equiv \hat{\mathbf{E}}^{(+)}(\mathbf{r})%
+\hat{\mathbf{E}}^{(-)}(\mathbf{r})}%
\nonumber\\%
&&=\sum_{s}\left(\frac{2\pi\hbar\omega_s}{{\cal V}}\right)^{1/2}%
\left[i\mathbf{e}_s a_s\mathrm{e}^{i\mathbf{k}_s\mathbf{r}}%
-i\mathbf{e}_s a_s^{\dagger}\mathrm{e}^{-i\mathbf{k}_s\mathbf{r}}\right]%
\nonumber\\%
&&=\hat{\mathbf{E}}_{\bot}(\mathbf{r})+\sum_{b=1}^{N}\frac{4\pi}{{\cal V}}%
\sum_{s}\mathbf{e}_s(\mathbf{e}_s\hat{\mathbf{d}}^{(b)})%
\mathrm{e}^{i\mathbf{k}_s(\mathbf{r}-\mathbf{r}_b)}%
\label{2.4}%
\end{eqnarray}%
Here $a_s$ and $a_s^{\dagger}$ are the annihilation and creation operators for
the $s$th field's mode and the quantization scheme includes the periodic
boundary conditions in the quantization volume ${\cal V}$. The bottom line in
Eq.(\ref{2.4}) indicates the important difference between the actual transverse
electric field denoted as $\hat{\mathbf{E}}_{\bot}(\mathbf{r})$ and the
displacement field. The difference cannot be ignored at the distances
comparable with either atomic size or the radiation wavelength, which is the subject of the present report. For such a dense configuration the
definitions (\ref{2.3}) and (\ref{2.4}) should be clearly understood.

Let us make a few remarks. The second term in Eq.(\ref{2.3}) reveals a
nonconverging self-energy (self-action) of the dipoles. This term is often
omitted in practical calculations since it does not principally affect the
dipoles' dynamics, particularly when the difference between transverse electric
and displacement fields is small. It can be also formally incorporated into the
internal Hamiltonian associated with the atomic dipoles. However, as was
pointed out in Ref. \cite{SKKH} via tracing the Heisenberg dynamics of atomic
variables, the self-action term is mostly compensated by the self-contact
dipole interaction. The latter manifests itself in the dipoles' dynamics when
$\mathbf{r}=\mathbf{r}_a=\mathbf{r}_b$ for interaction of a specific $a$-th
dipole in Hamiltonian (\ref{2.3}) with the longitudinal field created by the
same dipole in the second term in Eq. (\ref{2.4}). Both these nonconverging
self-action and self-contact interaction terms can be safely renormalized in
evaluation of a single-particle contribution into the self-energy part of the
perturbation theory expansion for the resolvent operator; see below.

\subsection{Resolvent operator and $N$-particle Green's function}\label{II.B}

The transition amplitude (\ref{2.1}) can be simplified if we substitute in it the interaction operator (\ref{2.3}) by keeping only the terms with annihilation of the incoming photon in the input state and creating the outgoing photon in the output state. Such a simplification is in accordance with the standard approach of the rotating wave approximation, which is surely fulfilled for a near-resonance scattering process. As a consequence of this approximation the transition amplitude is now determined by the complete resolvent operator projected onto the vacuum state for the field subsystem and onto the singly excited state for the atomic subsystem
\begin{equation}
\tilde{\hat{R}}(E)=\hat{P}\,\hat{R}(E)\,\hat{P}\equiv \hat{P}\frac{1}{E-\hat{H}}\hat{P}.%
\label{2.5}%
\end{equation}
Here we defined the following projector
\begin{eqnarray}
\lefteqn{\hspace{-0.8cm}\hat{P}=\sum_{a=1}^{N}\;\sum_{\{m_j\},j\neq a}\;\sum_{n}%
|m_1,\ldots,m_{a-1},n,m_{a+1},\ldots m_N\rangle}%
\nonumber\\%
&&\hspace{-0.5cm}\langle m_1,\ldots,m_{a-1},n,m_{a+1},\ldots,m_N|\times|0\rangle\langle 0|_{\mathrm{Field}}%
\label{2.6}%
\end{eqnarray}
which selects in the atomic Hilbert subspace the entire set of the states where
any $j$th of $N-1$ atoms populates a Zeeman sublevel $|m_j\rangle$ in its
ground state and one specific $a$th atom (with $a$ running from $1$ to $N$ and
$j\neq a$) populates a Zeeman sublevel $|n\rangle$ of its excited state. The
field subspace is projected onto its vacuum state and the operator
$\tilde{\hat{R}}(E)$ can be further considered as a matrix operator acting only
in atomic subspace. The elements of the $T$ matrix can be directly expressed by the
resolvent operator as follows:
\begin{eqnarray}
\lefteqn{T_{g'\mathbf{e}'\mathbf{k}',g\,\mathbf{e\,k}}(E)=\frac{2\pi\hbar\sqrt{\omega'\omega}}{{\cal V}}%
\sum_{b,a=1}^{N}\;\sum_{n',n}}%
\nonumber\\%
&&\hspace{1 cm}(\mathbf{d}\mathbf{e}')_{n'm'_b}^{*}(\mathbf{d}\mathbf{e})_{nm_a}%
\mathrm{e}^{-i\mathbf{k}'\mathbf{r}_b+i\mathbf{k}\mathbf{r}_a}%
\nonumber\\%
&&\langle\ldots m'_{b-1},n',m'_{b+1}\ldots |\tilde{\hat{R}}(E)%
|\ldots m_{a-1},n,m_{a+1}\ldots \rangle%
\nonumber\\%
&&\label{2.7}%
\end{eqnarray}
This performs a generalization of the well-known Kramers-Heisenberg formula
\cite{BerstLifshPitvsk} for the scattering of a photon by a many-particle system
consisting of atomic dipoles. The selected specific matrix element runs all the
possibilities when the incoming photon is absorbed by any $a$th atom and the
outgoing photon is emitted by any $b$th atom of the ensemble, including the
possible coincidence $a=b$. The initial atomic state is given by
$|g\rangle\equiv|m_1,\ldots,m_N\rangle$ and the final atomic state is given by
$|g'\rangle\equiv|m'_1,\ldots,m'_N\rangle$.

The projected resolvent operator contributing to Eq. (\ref{2.7}) is defined in the the
Hilbert subspace of a finite size with dimension $d_eN\,d_g^{N-1}$, where $d_e$
is the degeneracy of the atomic excited state and $d_g$ is the degeneracy of its
ground state. The matrix elements of operator $\tilde{\hat{R}}(E)$ can be
linked with the $N$-particle causal Green's function of atomic subsystem via
the following Laplace-type integral transformation:
\begin{eqnarray}
\lefteqn{\langle\ldots m'_{b-1},n',m'_{b+1}\ldots |\tilde{\hat{R}}(E)%
|\ldots m_{a-1},n,m_{a+1}\ldots \rangle}%
\nonumber\\%
&&\times\delta\left(\mathbf{r}'_1-\mathbf{r}_1\right)\ldots\delta\left(\mathbf{r}'_b-\mathbf{r}_b\right)\ldots%
\delta\left(\mathbf{r}'_a-\mathbf{r}_a\right)\ldots\delta\left(\mathbf{r}'_N-\mathbf{r}_N\right)%
\nonumber\\%
&&=-\frac{i}{\hbar}\int_0^{\infty}dt\,\,\exp\left[+\frac{i}{\hbar}E\,t\right]\,%
\nonumber\\%
&&G^{(N)}\left(1',t;\ldots ;b',t;\ldots ;N',t|1,0;\ldots ;a,t;\ldots;N,0\right)%
\label{2.8}%
\end{eqnarray}
where on the right side we denoted $j=m_j,\mathbf{r}_j$ (for $j\neq
a$) and $j'=m'_j,\mathbf{r}'_j$ (for $j'\neq b$), and for specific atoms
$a=n,\mathbf{r}_a$ and $b'=n',\mathbf{r}'_b$. Here
$\mathbf{r}_j=\mathbf{r}'_j$, for any $j=1\div N$, is the spatial location of
$j$th atom, which is assumed to be conserved in the scattering process. This
circumstance is expressed by the sequence of $/delta$ functions in Eq. (\ref{2.8}).
The causal Green's function is given by the vacuum expectation value of the
following chronologically ($T$)-ordered product of atomic second quantized
$\Psi$ operators introduced in the Heisenberg representation
\begin{eqnarray}
\lefteqn{\hspace{-0.5cm}G^{(N)}\left(1',t'_1;\ldots ;b',t'_b;\ldots ;N',t'_N|1,t_1;\ldots ;a,t_a;\ldots;N,t_N\right)}%
\nonumber\\%
&&=\langle T \Psi_{m'_1}(\mathbf{r}'_1,t'_1)\ldots\Psi_{n'}(\mathbf{r}'_b,t'_b)\ldots%
\Psi_{m'_N}(\mathbf{r}'_N,t'_N)%
\nonumber\\%
&&\Psi_{m_N}^{\dagger}(\mathbf{r}_N,t_N)\ldots%
\Psi_{n}^{\dagger}(\mathbf{r}_a,t_a)\ldots\Psi_{m_1}^{\dagger}(\mathbf{r}_1,t_1)\rangle,%
\nonumber\\%
\label{2.9}%
\end{eqnarray}
where $\Psi_{\ldots}(\ldots)$ and $\Psi_{\ldots}^{\dagger}(\ldots)$ are
respectively the annihilation and creation operators for an atom in a
particular state and coordinate. All the creation operators in this product
contribute to transform (\ref{2.8}) while being considered at initial time "$0$"
and all the annihilation operators are considered at a later time $t>0$. That allows us to ignore
effects of either bosonic or fermionic quantum statistics associated with
atomic subsystem as far as we neglect any possible overlap in atomic locations
and consider the atomic dipoles as classical objects randomly distributed in
space. We ordered operators in Eq. (\ref{2.9}) in such a way that in the fermionic
case (under the anticommutation rule) and without interaction it generates the
product of independent individual single-particle Green's functions associated
with each atom and with positive overall sign.

The perturbation theory expansion of the $N$-particle Green's function
(\ref{2.9}) can be visualized by the series of the diagrams in accordance with
the standard rules of the vacuum diagram technique; see Ref.
\cite{BerstLifshPitvsk}. After rearrangement the diagram expansion can be
transformed to the following generalized Dyson equation:
\begin{equation}
\scalebox{1.0}{\includegraphics*{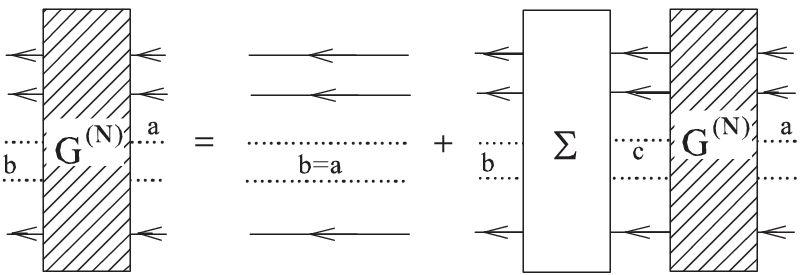}}%
\label{2.10}
\end{equation}
where the long straight lines with arrows correspond with individual causal
single-particle Green's functions of each atom in the ensemble such that the
first term on the right side performs the graph image of nondisturbed
$N$-particle propagator (\ref{2.9}). The dashed block edged by short
lines with arrows is the complete collective $N$-particle Green's function dressed by the
interaction. In each diagram block of equation (\ref{2.10}) we indicated by
$a,b,c$ (running from $1$ to $N$) the presence of one specific input as well as
an output line associated with the single excited state equally shared by all
the atoms of the ensemble. The sum of tight diagrams, which cannot be reduced to
the product of lower order contributions linked by
nondisturbed atomic propagators, builds a block of so-called self-energy part
$\Sigma$. The diagram equation (\ref{2.10}) in its analytical form performs an
integral equation for $G^{(N)}(\ldots)$. With its transformation to the energy
representation (\ref{2.8}) the integral equation can be recomposed to the set
of algebraic equations for the matrix of the projected resolvent operator
$\tilde{\hat{R}}(E)$, which can be further numerically solved. The crucial
requirement for this is the knowledge of the self-energy part (quasi-energy
operator acting in the atomic subspace), which as we show below can be
approximated by the lower orders in expansion of the perturbation theory.

\subsection{The self-energy part}

In the lower order of perturbation theory the self-energy part consists of two
contributions having single-particle and double-particle structures. Each
specific line in the graph equation (\ref{2.10}) associated with excitation of
an $a$th atom generates the following irreducible self-energy diagram:
\begin{eqnarray}
\lefteqn{\scalebox{1.0}{\includegraphics*{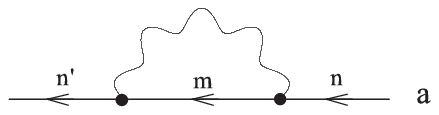}}}
\nonumber\\%
&&\Rightarrow\sum_{m}\int\frac{d\omega}{2\pi} d^{\mu}_{n'm}d^{\nu}_{mn}%
iD^{(E)}_{\mu\nu}(\mathbf{0},\omega)%
\nonumber\\%
&&\times\frac{1}{E-\hbar\omega-E_m+i0}%
\equiv\Sigma^{(a)}_{n'n}(E),%
\label{2.11}
\end{eqnarray}
which is analytically decoded with applying transformation (\ref{2.8}) in
the energy representation. Here the internal wavy line expresses the
causal-type vacuum Green's function of the chronologically ordered polarization
components of the field operators
\begin{equation}
iD^{(E)}_{\mu\nu}(\mathbf{R},\tau)=\left\langle T\hat{E}_{\mu}(\mathbf{r}',t')%
\hat{E}_{\nu}(\mathbf{r},t)\right\rangle,%
\label{2.12}%
\end{equation}
which depends only on difference of its arguments
$\mathbf{R}=\mathbf{r}'-\mathbf{r}$ and $\tau=t'-t$ and has the following
Fourier image:
\begin{eqnarray}
\lefteqn{D^{(E)}_{\mu\nu}(\mathbf{R},\omega)=\int_{-\infty}^{\infty} d\tau\,\mathrm{e}^{i\omega\tau}%
D^{(E)}_{\mu\nu}(\mathbf{R},\tau)}%
\nonumber\\%
&=&-\hbar\frac{|\omega|^3}{c^3}\left\{i\frac{2}{3}h^{(1)}_0\left(\frac{|\omega|}{c}R\right)\delta_{\mu\nu}\right.%
\nonumber\\%
&&\left.+\left[\frac{X_{\mu}X_{\nu}}{R^2}-\frac{1}{3}\delta_{\mu\nu}\right]%
ih^{(1)}_2\left(\frac{|\omega|}{c}R\right)\right\};%
\label{2.13}%
\end{eqnarray}
see Ref. \cite{BerstLifshPitvsk}. Here $h^{(1)}_L(\ldots)$ with $L=0,2$ are the
spherical Hankel functions of the first kind. As follows from Eq. (\ref{2.11})
the Green's function (\ref{2.13}) contributes in that expression in a
self-interacting form with spatial argument $\mathbf{R}\to\mathbf{0}$. As a
consequence the expression (\ref{2.11}) becomes non-converging in the limit
$R\to 0$ and the integration over $\omega$ is nonconverging. Part of
nonconverging terms should be associated with the longitudinal self-contact
interaction. These terms are compensated by the dipolar self-action; see
Eq. (\ref{2.3}) and the related remark given above. The residual nonconvergency
has radiative nature and demonstrates general incorrectness of the Lamb-shift
calculation in assumptions of the long-wavelength dipole approximation. Finally
we follow the standard renormalization rule,
\begin{eqnarray}
\Sigma^{(a)}_{n'n}(E)&=&\Sigma^{(a)}(E)\delta_{n'n},%
\nonumber\\%
\Sigma^{(a)}(E)&\approx&\Sigma^{(a)}(\hbar\omega_0)=\hbar\Delta_{\mathrm{L}}-i\hbar\frac{\gamma}{2},%
\label{2.14}%
\end{eqnarray}
where $\Delta_{\mathrm{L}}\to\infty$ is incorporated into the physical energy
of the atomic state. To introduce the single-atom natural decay rate $\gamma$
we applied the Wigner-Weiskopf pole approximation and substituted the
energy $E=\hbar\omega_k+E_g$ by its near resonance mean estimate $E\approx E_n$
with assumption that the atomic ground state is the zero-energy level such that
$E_g=\sum_{j=1}^{N}E_{m_j}=E_m=0$. Then energy of the excited state is given by
$E_n=\hbar\omega_0$, where $\omega_0$ is the transition frequency.

In the lower order of perturbation theory, the double-particle contribution to
the self-energy part consists of two complementary diagrams:
\begin{eqnarray}
\lefteqn{\scalebox{1.0}{\includegraphics*{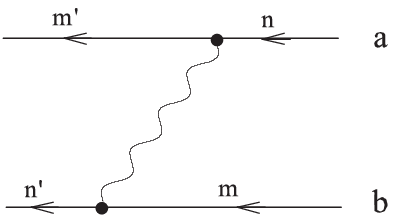}}}
\nonumber\\%
&&\hspace{-0.5cm}\Rightarrow\int\frac{d\omega}{2\pi} d^{\mu}_{n'm}d^{\nu}_{m'n}%
iD^{(E)}_{\mu\nu}(\mathbf{R}_{ab},\omega)%
\nonumber\\%
&&\hspace{-0.5cm}\times\frac{1}{E-\hbar\omega-E_m-E_{m'}+i0}%
\equiv\Sigma^{(ab+)}_{m'n';nm}(E)
\label{2.15}
\end{eqnarray}
and
\begin{eqnarray}
\lefteqn{\scalebox{1.0}{\includegraphics*{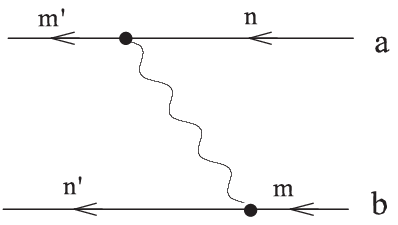}}}
\nonumber\\%
&&\hspace{-0.5cm}\Rightarrow\int\frac{d\omega}{2\pi} d^{\mu}_{n'm}d^{\nu}_{m'n}%
iD^{(E)}_{\mu\nu}(\mathbf{R}_{ab},\omega)%
\nonumber\\%
&&\frac{1}{E+\hbar\omega-E_n-E_{n'}+i0}%
\equiv\Sigma^{(ab-)}_{m'n';nm}(E),
\label{2.16}
\end{eqnarray}
which are responsible for the excitation transfer from atom $a$ to atom $b$
separated by a distance $R_{ab}$. The vector components of the dipole matrix
elements $d^{\nu}_{m'n}$ and $d^{\mu}_{n'm}$ are related with atoms $a$ and $b$
respectively. In the pole approximation $E\approx E_n=\hbar\omega_0$ the
$\delta$ function features dominate in the spectral integrals (\ref{2.15}) and
(\ref{2.16}) and the sum of both the terms gives
\begin{eqnarray}
\Sigma^{(ab)}_{m'n';nm}(E)&\approx&\Sigma^{(ab+)}_{m'n';nm}(\hbar\omega_0)+%
\Sigma^{(ab-)}_{m'n';nm}(\hbar\omega_0)%
\nonumber\\%
&=&\frac{1}{\hbar}\,d^{\mu}_{n'm}d^{\nu}_{m'n}\,D^{(E)}_{\mu\nu}(\mathbf{R}_{ab},\omega_0).%
\label{2.17}%
\end{eqnarray}
The derived expression has clear physical meaning. For nearly located atoms the real component of the double-particle contribution to the self-energy part reproduces the static interaction between two atomic dipoles. Its imaginary component is responsible for formation of cooperative dynamics of the excitation decay in the entire radiation process. For long distances, when the atomic dipoles are separated by the radiation zone, this term describes radiation interference between any pair of two distant atoms, which weakly reduces with the interatomic separation. For short distances or in a dense sample the cooperative effects become extremely important and the scattering process becomes strongly dependent on a particular atomic configuration.

It is a challenging problem to further improve the self-energy part by taking into consideration the higher orders of the perturbation theory expansion. Here we only substantiate the validity and sufficiency of the lower order approximation for the considered configuration. The main physical reason for this is weakness of interaction. This justifies ignoring of any deviation from free dynamics of atomic variables on a short time scale associated with the light retardation on distances of a few wavelengths. That yields main cooperation in the radiative dynamics among neighboring dipoles which can effectively interact via static longitudinal electric field. The diagram (\ref{2.16}), in contrast with (\ref{2.15}), is mostly important for evaluation of the static interaction such that in this graph the field propagator preferably links the points with coincident times on atomic lines. As a consequence, the presence of such diagram fragments as a part of any irreducible diagrams in higher orders would make the overall contribution small and negligible just because the static dipole-dipole interaction only weakly affects the dipoles' dynamics during the short retardation time, which can be roughly estimated by the wave period $2\pi/\omega_0$. For the same reason we can ignore any vertex-type corrections to the diagram (\ref{2.14}). Another part of the self-energy diagrams in higher orders can be associated with correction of the static interaction for itself. If the atomic system were as dense as the atoms were separated by a distance comparable with atomic size (much shorter than the radiation wavelength) then the description of the static interaction in the simplest dipole model would be inconsistent and insufficient. This correction is evidently ignorable for atomic ensemble with a density of a few atoms in a volume scaled by the cubic radiation wavelength. In this case the higher order static corrections are negligible as far as the dipole-dipole interaction is essentially less than the internal transition energy. As we can finally see, for the considered atomic systems, the self-energy part is correctly reproducible by the introduced lower order contributions.

\section{Results and discussion}

\subsection{Cooperative scattering from the system of two atoms}

\begin{figure}[t]\center
\includegraphics{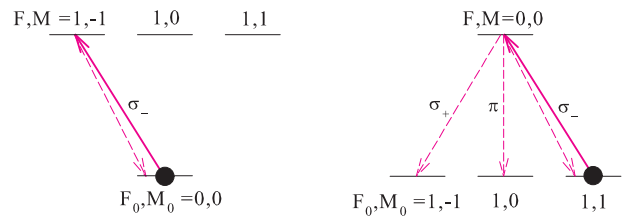}%
\caption{(Color online) The excitation diagram of "two-level" V-type atom (left) and overturned
"two-level" $\Lambda$-type atom (right). In both the configurations the light scattering is considered for
the left-handed $\sigma_{-}$ polarization mode. The $\Lambda$-atom populates the Zeeman
sublevel with the highest angular momentum projection.}
\label{fig1}%
\end{figure}

Let us apply the developed theory to the calculation of the total cross section
for the process of light scattering from the system consisting of two atoms. We
consider two complementary examples where the scattering atoms have different
but similar Zeeman state structure. In the first example we consider
V-type atoms, which have $F_0=0$ total angular momentum in the ground state
and $F=1$ total angular momentum in the excited state. Such atoms are the
standard objects for discussion of the Dicke problem see Ref. \cite{BEMST}, and
each atom performs a "two-level" energy system sensitive to the
vector properties of light. In an alternative example we consider the
$\Lambda$-type atoms, which can be also understood as overturned "two-level"
system, which have $F_0=1$ total angular momentum in the ground state and $F=0$
total angular momentum in the excited state. For the latter example in the
scattering scenario we assume the initial population by atoms of a particular
Zeeman sublevel of the ground state, which has highest projection of the
angular momentum. Both the excitation schemes and transition diagrams in the
laboratory reference frame are displayed in Fig. \ref{fig1}.

\begin{figure}[t]\center
\includegraphics{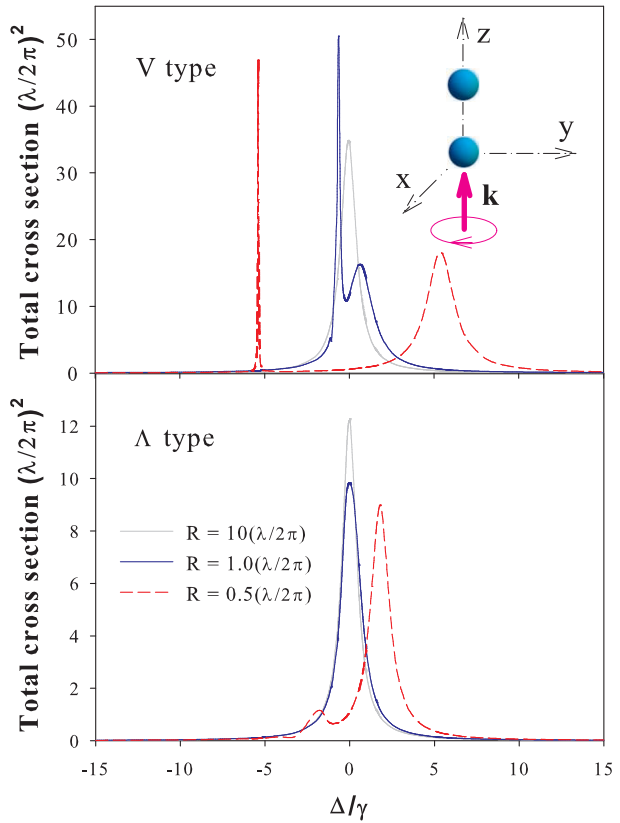}%
\caption{(Color online) Spectral dependencies of the total cross section for a photon
scattering from the system of two "two-level" V-type atoms (upper panel) and
$\Lambda$-type atoms (lower panel) in the collinear excitation geometry; see inset. In the case of V-type
atoms, in accordance with predictions of the Dicke model \cite{ChTnDRGr,BEMST}, the observed
resonances demonstrate either super- or subradiant behavior when interatomic
separation $R$ becomes shorter. In the case of $\Lambda$-type atoms the resonances are
less resolved and both have a line width comparable with atomic natural decay rate.}
\label{fig2}%
\end{figure}

\begin{figure}[t]\center
\includegraphics{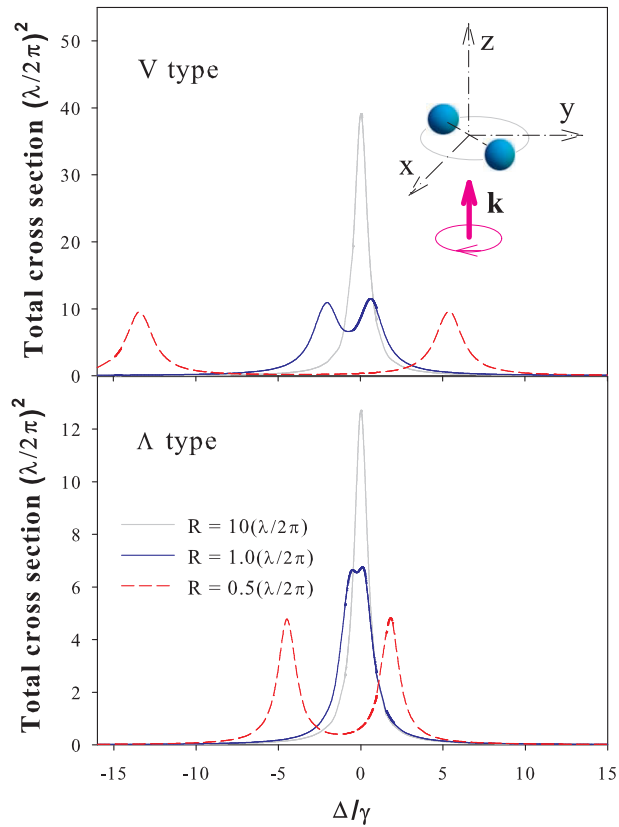}%
\caption{(Color online) Same as in Fig. \ref{fig2} but for planar excitation
geometry. In both the excitation schemes for either V- or $\Lambda$-type atoms there is a symmetric
resonance structure; see the text.}
\label{fig3}%
\end{figure}

\begin{figure}[t]\center
\includegraphics{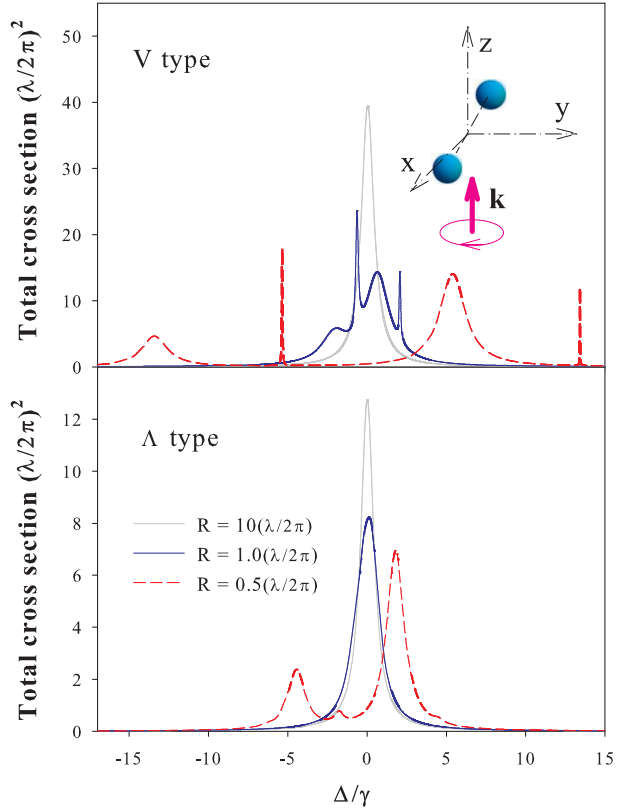}%
\caption{(Color online) Same as in Fig. \ref{fig2} but for random excitation
geometry. For V-type atoms there are two superradiant and two subradiant resonances.
For $\Lambda$-type atoms the four resonances are
less resolved and have line widths comparable with a single-atom natural decay rate.}
\label{fig4}%
\end{figure}

In Figs. \ref{fig2}-\ref{fig4} we reproduce the spectral dependencies of the
total cross section for a photon scattering from the system consisted of two
atoms separated by different distances $R$ and for different spatial
orientations. The variation of interatomic separation from $R=10\lambdabar$
(independent scatterers) to $R=0.5\lambdabar$ (strongly dependent scatterers)
transforms the scattering process from its independent to cooperative dynamics.
In the plotted graphs the frequency spectra, reproduced as function of the
frequency detuning $\Delta=\omega-\omega_0$ of the probe frequency $\omega$
from the nondisturbed atomic resonance $\omega_0$, are scaled by the natural
radiation decay rate of a single atom $\gamma$, which is significantly
different for $\Lambda$- and V-type energy configurations, such that
$\gamma(\Lambda)=3\gamma(V)$. As a consequence the near-field effects
responsible for the resonance structure of the resolvent operator and the
cross section manifest more perceptibly for the V-type atoms,
which are traditionally considered in many discussions of the Dicke system in
literature. In the symmetric collinear excitation geometry, when the internal
reference frame coincides with the laboratory frame see Fig. \ref{fig2} the
left-handed $\sigma_{-}$ excitation channel shown in Fig. \ref{fig1} is only
allowed for either V- or $\Lambda$-type transition schemes. In such a
symmetric configuration the interatomic interaction via the longitudinal as
well as via the radiative transverse fields splits the excitation transition in
two resonance lines. For the case of the V-type excitation the observed
resonances demonstrate either superradiant or subradiant nature. This is an
evident indicator of the well-known Dicke effect of either cooperative or
anticooperative contribution of the atomic dipoles into the entire radiation
and scattering processes; see Refs. \cite{ChTnDRGr,BEMST}. For the system of two
$\Lambda$-type atoms separated by the same distances the atomic line also
splits in two resonances, but they are less resolved and have relatively
comparable line widths. The spectral widths indicate a slight cooperative
modification, which is much weaker effect than in the case of V-type atoms.
The physical reason of that is the contribution of the Raman scattering
channels, which are insensitive to the effects of dependent scattering.

If both the atoms are located in the wavefront plane of the driving field, as shown in Fig. \ref{fig3}, the spectral dependence of the cross section
is also described by two resonance features. With referring to the excitation
scheme defined in the laboratory frame see Fig. \ref{fig1} in the specific
planar geometry the double-particle self-energy part (\ref{2.17}) can couple
only the states $|1,\pm 1\rangle$ related to either upper (V-type) or lower
($\Lambda$-type) atomic levels. As a consequence the resolvent operator
$\tilde{\hat{R}}(E)$ has a block structure and only its 4 $\times$ 4
block, built in subspace $|0,0\rangle_1|1,\pm 1\rangle_2,\, |1,\pm
1\rangle_1|0,0\rangle_2$, can actually contribute to the scattering process. We
subscribed the states by the atomic number $a,b=1,2$. The eigenstates of this
matrix have different parities $g$ (even) and $u$ (odd) reflecting their
symmetry or antisymmetry to transposition of the atomic state; see Ref.
\cite{LaLfIII}.\footnote{By "parity" we mean symmetry of the self-energy
part to the transposition of atoms. This is similar to the parity definition
for homonuclear diatomic molecules in chemistry; see Ref. \cite{LaLfIII}}. The
observed resonances can be associated with two even-parity states symmetrically
sharing the single excitation in the system of two atoms. Such selection rule
is a consequence of the evident configuration symmetry of the system, shown in
inset of Fig. \ref{fig3}, to its rotation on any angle around the $z$ axis such
that the allowed transition amplitude should be insensitive to the atoms'
positions. In contrast to the collinear geometry case in the planar geometry
both the resonances have identical shapes and line widths. It is also
interesting that for this specific excitation geometry both the atomic systems
of either V- or $\Lambda$-type demonstrate similar spectral behavior.

In general for random orientation of the diatomic system shown in Fig.
\ref{fig4} there are four resonances. These resonances can be naturally
specified in the internal reference frame, where the quantization axis is directed
along the internuclear axis, via following the standard definitions of
diatomic molecule terms; see Ref. \cite{LaLfIII}. There are two $\Sigma_g$ and
$\Sigma_u$ terms of different parity and two doubly degenerate $\Pi_g$ and
$\Pi_u$ terms, which also have different parities. Here the defined terms are
associated with the symmetry of the self-energy part and specified by the
transition type in the internal frame such that the transition dipole moment
can have either $0$ projection ($\Sigma$ term) or $\pm 1$ projection
($\Pi$ term). For random orientation all these resonances can be excited and in
case of the V-type atoms the odd-parity resonances have subradiant nature
and the even-parity ones are superradiant. In contrast in the case of the
$\Lambda$-type atoms the observed resonances are less resolved and have comparable
widths; two of them have rather small amplitudes (see the lower panel of Fig.
\ref{fig4}). The previous configurations with the collinear and planar
excitation geometries respectively correspond to the excitations of the $\Pi_g$
and $\Pi_u$, and $\Sigma_g$ and $\Pi_g$, resonance pairs. Summarizing the
results, we can point out that all the plotted dependencies demonstrate
significant difference in the cooperative scattering dynamics resulting from
the similar quantum systems shown in Fig. \ref{fig1}.

\subsection{Cooperative scattering from a collection of $\Lambda$-type
atoms randomly distributed in space}

Evaluation of the resolvent operator for the situation of a many-particle system
is a challenging task and its solution depends on the type of
transition driven in the atomic system. For V-type atoms the problem can be
solved even for a macroscopic atomic ensemble since the number of equations rises
linearly with the number of atoms; see the relevant estimate given in Sec.
\ref{II.B}. In Ref. \cite{SKKH} the transformation of light scattering on
macroscopic atomic ensemble consisting of V-type atoms were analyzed as
functions of the sample density. Particularly, the authors demonstrated how the smooth
spectral dependence of the cross section, observed in the limit of dilute and
weakly disordered distribution of atomic scatterers, would transform to the
random speckle resonance structure in the case of strongly disordered and dense
distribution containing more than one atom in the cubic wavelength. The
presence of narrow sub-radiant Dicke-type resonance modes revealed a
microcavity structure built up in an environment of randomly distributed atomic
scatterers that can be posed as a certain analog of Anderson-type localization
of light.

Our analysis in the previous section indicates that in the example of
the $\Lambda$-type atoms the subradiant modes are not manifestable and such a
system would be not suitable for observation of the localization effects.  For
coherent mechanisms of the quantum memory, which we keep in mind as a most
interesting implication, the existence of the localization regime would be
useful but not a crucially important feature of the light propagation process.
However, the spectral profile of the scattering cross section and its
dependence on the atomic density and sensitivity to the level of disorder are
very important, for example, for further consideration of an EIT-based memory
scheme.

Below we consider an example of the atomic system consisted of five
$\Lambda$-type atoms, which is described by the $405\times 405$ square matrix
of the resolvent operator $\tilde{R}(E)$. With evident provisoes but at least
qualitatively the system can be considered as having many particles and can show a
tendency toward macroscopic behavior. We show how the scattering process is
modified when the configuration is made more dense and how this corresponds
with the description of the problem in terms of the macroscopic Maxwell theory.
In macroscopic description the atomic system can be approximated by a
homogeneous dielectric sphere of a small radius, which scatters light in
accordance with the Rayleigh mechanism; see Ref. \cite{BornWolf}. We fix the
parameters of the dielectric sphere by the same density of atoms as we have in
the compared microscopic random distribution. The calculation of the dielectric
susceptibility were made similarly to that done earlier in
Ref. \cite{SKKH} and we will publish the calculation details elsewhere. The key point of our numerical analysis is to verify the presence
of the Zeeman structure, which manifests itself via the Raman scattering
channels in the observed total scattering cross section.

\begin{figure}[t]\center
\scalebox{0.9}{\includegraphics{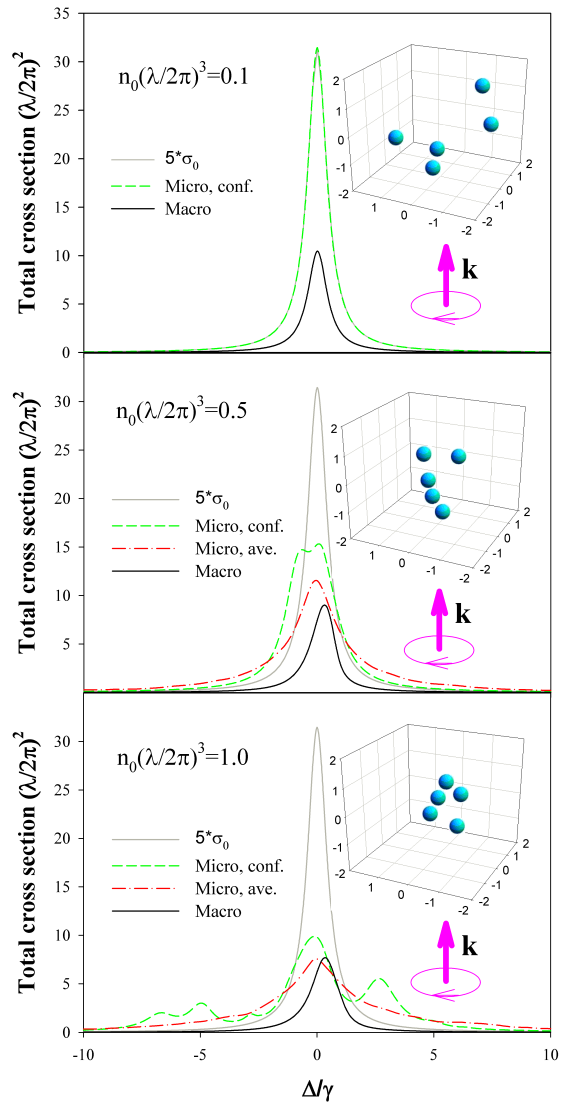}}%
\caption{(Color online) Spectral dependencies of the total cross section for a photon
scattering from the system of five $\Lambda$-type atoms randomly distributed in the space
for different densities $n_0\lambdabar^3=0.1,\,0.5,\,1.0$. The solid gray curve indicates
the spectral profile of the cross section for five independent scatterers. The black solid curve
performs the calculations for Rayleigh particle, which are based on macroscopic Maxwell theory.
The green [light gray] dashed curves show the microscopic calculations done for a particular configuration and
the red [dark gray] dashed-dotted curves show the statistical averaging over many configurations.}
\label{fig5}%
\end{figure}

In Fig. \ref{fig5} we show how the scattering cross section is modified with
varying atomic density $n_0$, scaled by the light bar wavelength $\lambdabar$, from
$n_0\lambdabar^3=0.1$ (dilute configuration) to $n_0\lambdabar^3=1$ (dense
configuration). There are two reference dependencies shown in these plots and
indicated by dashed and solid black curves. The dashed curve is the spectral
profile of single-atom cross section $\sigma_0=\sigma_0(\Delta)$ multiplied by
the number of atomic scatterers $N=5$. The solid black curve is evaluated via the
self-consistent macroscopic Maxwell description and reproduces the scattering
cross section for the Rayleigh particle performed by a small dielectric sphere.
Other dependencies subsequently show the results of microscopic calculations of
the scattering cross section: (green [dashed light gray]) for a particular random configuration
(visualized in insets) and (red [dash-dotted dark gray]) the microscopic spectral profiles averaged over
many random configurations.

The upper panel of Fig. \ref{fig5} relates to the low-density (i.e., dilute configuration or weak disorder) regime, which is insensitive to any specific location of atomic scatterers in space. Indeed the exact result evaluated with the microscopic model is perfectly reproducible by the simplest approximation of the cross section by the sum of partial contributions of all five atoms considered as independent scatterers. This confirms the traditional vision of light propagation through a multiparticle atomic ensemble as through the system of independent scatterers, which are in background of many practical scenarios of interaction of atomic systems with external fields. The Raman channel manifests in the scattering process, a direct consequence of the Zeeman degeneracy of the atomic ground state. In contrast, the central and bottom panels of Fig. \ref{fig5} show how the scattering process is modified in the situation of high density and strong disorder when the near-field effects are manifestable. The system evidently demonstrates cooperative behavior and the scattering mechanism becomes extremely sensitive to any specific distribution of the scatterers in space. The spectral profile is described by several resonances, and locations, amplitudes, and widths are unique for each specific configuration. However, there is a certain tendency to compromise the microscopically calculated scattering profile with the rough macroscopic prediction. The latter keeps only the Rayleigh channel as observable in the self-consistent macroscopic model of the scattering process. It is interesting that for any configurations, created randomly in the spatial distribution of the atomic scatterers, one of the observed resonances is preferably located near the vicinity of the zero detuning $\Delta\sim 0$. As a consequence, after the configuration averaging, the system demonstrates scattering characteristics qualitatively similar to those reproduced by the macroscopic model.

\subsection{Application to atomic memory problem}

The considered system of $\Lambda$-type atoms has certain potential for light-assisted coherent redistribution of atoms in the ground state initiated by simultaneous action of strong control and weak signal modes, that is, for realization of atomic memory protocol. Let us discuss the applicability and advantage of such a dense configuration of atoms for realization of light storage in atomic memories. At present most of the experiments and the  supporting theoretical discussions operate with dilute configurations of atoms either confined with MOT at low temperature or existing in a warm vapor phase; see Refs. \cite{PSH,Simon,SpecialIssueJPhysB}. For such systems the standard conditions for realization of either EIT- or Raman-based storage schemes require an optical depth of around hundreds such that the macroscopic ensemble would typically consist of billions of atoms. The optimization of the memory protocol for the parameters of optical depth, pulse shape, etc., has been the subject of many discussions in literature, see Ref. \cite{NGPSLW} and references therein. There would be an evident advantage in developing the memory unit with fewer atoms but with the same optical depth of the sample. This immediately readdresses the basic problem of cooperative light scattering by a dense system of the $\Lambda$-type configured atoms.

The presented microscopic analysis of the scattering process in such systems shows that in the strong disorder regime the spectral profile of the cross section is generally described by rather complicated and randomized resonance structure contributed by both the longitudinal and transverse self-energy interaction parts of the resolvent operator. This spectrum is unique for each particular configuration of the atomic scatterers and has only slight signature of original nondisturbed atomic spectrum. This circumstance is a direct consequence of the complicated cooperative dynamics, which reflects a microcavity nature of light interaction with a strongly disordered atomic ensemble.

 To determine possible implications of our results to the problem of atomic memories, we should extend the presented calculations toward the ensembles consisting of a macroscopic number of atoms. Such an extension seems not so straightforward since the number of contributing equations rises exponentially with the number of atoms and certain simplifying approximations are evidently needed. In this sense our calculations of the scattering cross section performed for a small collection of atoms can be considered a precursor to calculation of the transmittance coefficient, which would be a key characteristic in the macroscopic description of the problem. Our calculations indicate preferable contribution of the Rayleigh mechanism in the overall cooperative scattering process for a density and disorder level near the Ioffe-Regel bound $n_0\lambdabar^3\sim 1$. It is important that in this case one of the absorption resonances is located in the spectral domain near the zero detuning for any atomic configurations and provides the desirable conditions for further observation of the EIT phenomenon. The presence of the control mode, tuned at this predictable resonance point and applied in any "empty" arm of the $\Lambda$ scheme see Fig. \ref{fig1}, would make the atomic sample transparent for a signal pulse. Due to controllable spectral dispersion the signal pulse could be delayed and effectively converted into the long-lived spin coherence.

Realization of this scheme requires essentially fewer atoms than for dilute ensembles prepared in warm vapors and in MOT experiments. Roughly for a fixed optical depth $b_0\sim n_0\lambdabar^2L$, where $L$ is the sample length, and for $n_0\lambdabar^3\ll 1$, the required number of atoms, allowing for diffraction losses, should be more than $b_0^2/n_0\lambdabar^3$. This number can be minimized if we approach the dense configuration $n_0\lambdabar^3\sim 1$ and make the near field effects manifestable. We are currently working on a self-consistent modification of the presented calculation scheme to make it applicable for a multiatomic ensemble and then to describe the problem in a macroscopic limit. This can be done if we take into consideration the near-field effects only for the neighboring atoms separated by a distance of wavelength. For the intermediate densities with $n_0\lambdabar^3\sim 1$ we can soften our original estimate, given in Sec. \ref{II.B}, for the number of equations to be solved, and can expect that the actual number would be scaled as $d_eN\,d_g^{n-1}$. Here $n-1\sim n_0\lambdabar^3$ performs the varying parameter denoting the number of the neighboring atoms, which have near field interfering with a selected specific atom.  Our preliminary analysis shows that such a calculation algorithm should demonstrate a rapidly converging series with increasing $n$ and would allow us to include the control mode in the entire calculation procedure. Such a modification of the performed calculation scheme would be practically important and generally interesting for better understanding the microscopic nature of a $\Lambda$-type optical interaction in macroscopic atomic systems existing in a strong disorder regime.

\section{Summary}
In this paper we have studied the problem of light scattering on a collection of atoms with degenerate structure of the ground state, which cooperatively interact with the scattered light. We have discussed the difference for the scattering process between such system of atoms and well-known object of the Dicke problem, performing an ensemble of two-level V-type atoms. The investigation is specifically focused toward understanding principle aspects of the scattering processes that can occur and how they vary as the atomic density is varied from low values to levels where the mean separation between atoms is on the order of the radiation wavelength. For both the $\Lambda$- and V-type systems the spectral profile of the scattering cross section strongly depends on the particular atomic spatial configuration. However, in the case of the degenerate ground state, the presence of Raman scattering channels washes out visible signature of the super- and subradiant excitation modes in the resolvent spectrum, which are normally resolved in the system consisted of two-level atoms. We have discussed advantages of the considered system in the context of its possible implications for the problem of light storage in a macroscopic ensemble of dense and ultracold atoms and we point out that the quantum memory protocol can be effectively organized with essentially fewer atoms than in the dilute configuration regime.

\section*{ACKNOWLEDGMENTS}

We thank Elisabeth Giacobino, Igor Sokolov, Ivan Sokolov, and Julien Laurat for fruitful discussions. The work was supported by RFBR 10-02-00103, by the CNRS-RFBR collaboration (CNRS 6054 and RFBR 12-02-91056) and by Federal Program "Scientific and Scientific-Pedagogical Personnel of Innovative Russia on 2009-2013" (Contract No. 14.740.11.1173 ).

\end{document}